\documentstyle[12pt,a4]{article}

\def\a{\alpha}
\def\b{\beta}
\def\d{\delta}
\def\e{\epsilon}
\def\s{\sigma}
\def\h{\sqrt{h}}
\def\hh{\hat{h}_{ab}}
\def\labelenumi{(\theenumi)}
\def\theenumi{\roman{enumi}}

\begin{document}

\rightline{YITP-96-41}
\rightline{gr-qc/9609052}

\vskip 1.5 cm 
\centerline{\Large 
Partition Function for (2+1)-Dimensional Einstein Gravity}
\vskip 1.0 cm 
\centerline{\large {\sc Masafumi Seriu}}
\centerline{\it  Yukawa Institute for Theoretical Physics}
\centerline{\it   Kyoto University, Kyoto 606, Japan}
\centerline{and}
\centerline{\it Institute of Cosmology, 
                   Physics \& Astronomy Department}
\centerline{\it Tufts University, Medford, 
                      Massachusetts 02155, USA}

\vskip 1.5cm

\begin{abstract}
 Taking   (2+1)-dimensional pure Einstein gravity
 for  arbitrary genus $g\geq 1$ as a model,  
 we investigate the  relation between the   partition function 
 formally defined on the entire phase space and 
 the one written in terms of the reduced phase space.   
 The  case of $g=1$ (torus) is analyzed in detail and it 
 provides us with good lessons for  quantum cosmology.
 
  We  formulate the gauge-fixing conditions 
 in a form suitable for our purpose. 
 Then the gauge-fixing procedure is applied to  
 the partition function $Z$ for  (2+1)-dimensional  gravity,
 formally defined on the entire  phase space. 
  We show that 
  basically it reduces to a partition function defined 
 for the reduced system, 
 whose dynamical variables are $(\tau^A, p_A)$. [Here the 
 $\tau^A$'s are the Teichm\"uller parameters, and the $p_A$'s 
 are their conjugate momenta.] 
 
As for the case of $g=1$, we find out that  $Z$ is also related 
with   another reduced form,  
whose dynamical variables 
are not only $(\tau^A, p_A)$, but also $(V, \s)$. [Here $\s$ is 
a conjugate momentum to the 2-volume (area) $V$ of a spatial 
section.] 
A  nontrivial factor appears in the 
measure  in terms of this type of
 reduced form. This factor is understood as a Faddeev-Popov 
 determinant associated with the time-reparameterization 
 invariance inherent in this type of formulation.
  In this manner, the relation between two reduced formulations 
  becomes transparent in the context  of quantum theory.

 As another result  for the case of $g=1$,  one factor 
 originating from the zero-modes of a differential operator 
 $P_1$ can appear in the path-integral measure in the reduced 
 representation of $Z$. It depends on how to define 
 the path-integral domain for the shift vectors $N_a$ in $Z$: 
If it is defined to include $\ker P_1$, the nontrivial factor 
does not  appear. On the other hand, if the integral domain is 
defined to exclude $\ker P_1$, the factor appears in the measure. 
 This factor can depend on the dynamical 
 variables, typically as  a function of  $V$, and 
  can  influence  the semiclassical dynamics of the 
  (2+1)-dimensional spacetime.

  These results shall be 
  significant from the viewpoint of quantum gravity and quantum 
  cosmology. 
   
\end{abstract}


\section{Introduction}
\label{section:I}

 Because both of its simplicity and nontrivial nature,
  (2+1)-dimensional Einstein gravity serves as a good 
  test case for pursuing  quantum gravity in 
  the framework of general relativity.  In particular, 
  because of the low dimensionality, 
  the global degrees of freedom of a space can be analyzed 
  quite explicitly in this case
  ~\cite{MART,MONC,HOS,CAR1}.

  Recently,  back-reaction effects from quantum matter 
  on the global degrees of freedom of a semiclassical universe 
  were  analyzed explicitly~\cite{MS1}. In this analysis,   the 
  (2+1)-dimensional homogeneous spacetime with topology
  ${\cal M}\simeq T^2 \times {\bf R}$ was chosen as a model. 
  This problem was  investigated from a general interest on 
  the global properties of a semiclassical universe, whose 
  analysis has  not yet been pursued  
  sufficiently~\cite{MS1,MS2,MS3}.

  In this analysis, it was also investigated  whether 
   the path-integral measure could give a correction 
   to  the semiclassical dynamics of
  the global degrees of freedom~\cite{MS1}. 
  By virtue of several  techniques developed in string theory, 
  one can give a meaning to  a partition function, formally 
  defined as 
\begin{equation}
 Z={\cal N} \int [dh_{ab}][d\pi^{ab}][dN][dN_a]\  \exp iS\ \ \ .
\label{eq:partition0} 
\end{equation} 
Here $h_{ab}$ and $\pi^{ab}$ are a spatial metric and its 
conjugate momentum, respectively; $N$ and $N_a$ are the lapse 
function and the shift vector, respectively; 
$S$ is the canonical action 
for Einstein gravity.
 It  is expected that $Z$ reduces to   the form
\[
Z={\cal N} 
 \int [dV\ d\s][d\tau^A\ dp_A][dN'] \ 
  \mu(V, \s, \tau^A, p_A) \ 
   \exp i S_{ reduced}\ \ \ .
\]
Here $V$, $\s$, $\tau^A$, and  $p_A$ ($A=1,2$) are, respectively,
 the 2-volume (area) of a torus, its conjugate momentum, 
 the Teichm\"uller parameters, and their conjugate momenta;  
 $N'$ is the spatially constant part of $N$; 
 $S_{ reduced}$ is the reduced action written in terms of 
  $V$, $\s$, $\tau^A$ and $p_A$. The factor 
  $\mu(V, \s, \tau^A, p_A)$ is a possible nontrivial measure, 
  which can cause a modification of the semiclassical evolution 
  determined  by $S_{ reduced}$. The  
   result of Ref.\cite{MS1} was that 
  $\mu(V, \s, \tau^A, p_A)=1$: The partition function defined 
  as in  Eq.(\ref{eq:partition0}) is equivalent, after 
  a suitable gauge fixing, 
   to the one defined  directly from the reduced system, 
   $S_{ reduced}$. 
   Though this result looks natural at first sight, 
   it is far from trivial. 
   One needs to extract a finite dimensional reduced phase space 
   from an infinite dimensional original phase space. Therefore, 
   it is meaningful to show that such a natural reduction 
   is really achieved by a suitable gauge fixing.

  The main interest in Ref.\cite{MS1} was the explicit analysis 
  of the semiclassical dynamics of a tractable model, 
  ${\cal M}\simeq T^2 \times {\bf R}$. 
  Therefore,   the analysis of  
  the reduction of the partition function    
  was inevitably restricted to 
    the special model in question. Namely   
  it was the  case of $g=1$, where $g$ is a genus of a Riemann 
  surface. Furthermore, the model  was set to be spatially 
  homogeneous from the outset. 
  It is then desirable for completeness  to generalize the 
  analysis  in Ref.\cite{MS1} to the general  case of  any 
  $g\geq 1$.

  More significantly, there is  one issue remaining to be 
  clarified in the case of $g=1$: The relation  
  between the reduced system of the 
  type of Ref.\cite{MONC} and the one of the type of 
  Ref.\cite{HOS} in the context of  quantum theory.
   For brevity, let us call the former formulation as the 
   $\tau$-form, while the latter one as the 
   $(\tau, V)$-form. 
   The $\tau$-form takes $(\tau^A, p_A)$ as fundamental 
   canonical pairs and the action is given by~\cite{MONC}
\begin{equation}
S[\tau^A, p_A]
=\int d\s \{ p_A\  d{\tau}^A / d\s- V(\s, \tau^A, p_A) \} \ \ \ .
\label{eq:tau}
\end{equation}    
   On the other hand the $(\tau, V)$-form 
   uses $(V, \s)$ as well 
   as $(\tau^A, p_A)$ and the action is given in the 
   form~\cite{HOS}
\begin{equation}
S[(\tau^A, p_A), (V, \s)]
=\int dt\  \{ p_A\  \dot{\tau}^A + \s \dot{V}- N\ 
 H(\tau^A, p_A, V, \s) \}\ \ \ .
\label{eq:tauV}
\end{equation}    
[The explicit expression for $H$ shall be presented later 
(Eq.(\ref{eq:reducedaction})).]
 The key procedure in deriving  the 
  $(\tau, V)$-form (in the classical sense) 
  is to choose $N=$ spatially constant~\cite{HOS}.  
  Since the compatibility of this choice with 
   York's time-slicing 
  is shown by means of the equations of motion~\cite{HOS}, 
  one should investigate the effect of this choice 
   in quantum theory.
    Furthermore, the condition $N=$ spatially constant 
    is not in the standard form of the canonical gauge, so that 
    the analysis of its role in the quantum level 
    requires special cares.  
     Since the model analyzed in Ref.\cite{MS1} was 
    chosen to be spatially homogeneous, 
    this issue did not make its appearance. 
    We shall make these issues clarified. 

   Regarding the $(\tau, V)$-form, there is 
 another issue which is not very clear.  
  In this formulation, $(V, \s)$ joins to 
  $(\tau^A, p_A)$ as one of the  canonical pairs. 
  Therefore~\cite{MS1}, $\int[d\s]$
  should  appear in the final form of $Z$ 
  as well as $\int[dV]$.
  Since the adopted gauge-fixing condition is $\pi/\h -\s=0$  
   (York's gauge~\cite{YORK}), $\s$ plays the role of 
    a label parameterizing a family of allowed gauge-fixing 
    conditions, 
    so that  it is not dynamical in the beginning.
    Therefore, the appearance of $\int[d\s]$ is not  apparent, 
    and  worth to be traced  from a viewpoint of 
    a general procedure of  gauge fixing. 
    We shall investigate these points.

  Independently from the analysis of Ref.\cite{MS1}, Carlip also 
  investigated  the relation between two partition 
  functions, 
  one being defined on  the entire phase space, and the other 
  one on
   the reduced phase space in the sense of the 
   $\tau$-form~\cite{CAR2}. 
   With regard to   this problem,  
   his viewpoint was   more general 
   than Ref.\cite{MS1}. He showed that, for the case of 
   $g\geq 2$, 
   the partition function formally defined as 
   in Eq.(\ref{eq:partition0})
    is equivalent to the one for the reduced system in the 
    $\tau$-form. On the other hand, the exceptional case of 
    $g=1$ was not analyzed so much. 
    Indeed, we shall see later that the case of $g=1$ can yield a 
    different result compared with the case of $g \geq 2$. 
    In this respect, his analysis and the analysis 
    in Ref.\cite{MS1}
    do  supplement  each other. Furthermore, 
    his way of  analyzing is quite different 
    from the one developed in Ref.\cite{MS1}. 
    In particular, it looks difficult to trace 
    the appearance of $\int[d\s]$ if his analysis 
    is applied to the case of $g=1$ in the 
    $(\tau, V)$-form. It may be useful, therefore, 
     to investigate  all  the cases of $g\geq 1$ from a
      different angle, 
     namely by a developed version of the method of 
     Ref.\cite{MS1}.

  In view of these situations of previous work, we shall 
   present here 
  the full analysis for all the cases $g\geq 1$.
  In particular, a more detailed investigation for
    the case of $g=1$
 shall be performed.

  In \S \ref{section:II}, we shall investigate for $g\geq 1$ 
the reduction of the 
partition function of  Eq.(\ref{eq:partition0}), to the one for 
the reduced system in the $\tau$-form.
In \S \ref{section:III}, 
we shall investigate how the $(\tau, V)$-form  
emerges for $g=1$ in the course of 
the reduction of the partition function, 
Eq.(\ref{eq:partition0}). We shall find out that 
a nontrivial measure appears in the formula defining 
a partition function, if the $(\tau, V)$-form is 
adopted. We shall see that this factor is understood as 
the Faddeev-Popov determinant associated with the 
reparameterization invariance inherent in the $(\tau, V)$-form. 
Furthermore we shall see that another factor can appear in the 
measure for the case of $g=1$, originating from the 
existence of the zero modes of a certain differential operator
 $P_1$. It depends on how to define 
 the path-integral domain for the shift vector $N_a$ in $Z$: 
If it is defined to include $\ker P_1$, the nontrivial factor 
does not  appear, while it appears  if the integral domain is 
defined to exclude $\ker P_1$. 
We shall  discuss that this factor 
can influence the semiclassical dynamics of the 
(2+1)-dimensional spacetime with $g=1$.
 These observations urge us to clarify how to choose 
 the integral domain for $N_a$ in quantum gravity. 
 Section \ref{section:IV}  is devoted to several discussions.
In {\it Appendix}, 
we shall derive  useful formulas which shall become  
 indispensable for our analysis.


 \section{The partition function for (2+1)-gravity}
 \label{section:II}
 Let us consider a    (2+1)-dimensional spacetime, 
 ${\cal M}\simeq \Sigma \times {\bf R}$, where 
 $\Sigma$ stands for a compact, closed, orientable 
 2-surface with genus $g$. The partition function for 
 (2+1)-dimensional pure Einstein gravity is formally given by 
 \begin{equation}
 Z={\cal N} \int [dh_{ab}][d\pi^{ab}][dN][dN_a]
   \exp i\int dt \int_{\Sigma} d^2x \  (\pi^{ab} \dot{h}_{ab} 
   -N{\cal H} -N_a {\cal H}^a), \ \ \ 
 \label{eq:partition}
 \end{equation}
 where \footnote{
      We have 
      chosen  units such that  $c=\hbar=1$ and such that the 
      Einstein-Hilbert 
      action becomes just $\int R \sqrt{-g}$ up to a boundary 
      term.
       The spatial indices $a,b, \cdots$ are raised and lowered 
       by $h_{ab}$. The operator $D_a$ is the covariant 
       derivative 
       w.r.t. (with respect to) 
       $h_{ab}$, and  $^{(2)}  \! R$ stands for a scalar 
       curvature
        of the 2-surface $\Sigma$. Unless otherwise stated, 
        the symbols $\pi$ and $h$ stand
         for $h_{ab}\pi^{ab}$ and $\det h_{ab}$, respectively, 
          throughout this  paper.}
\begin{eqnarray}
{\cal H} &=& (\pi^{ab}\pi_{ab} - \pi^2)\sqrt{h^{-1} } - 
( ^{(2)}  \! R - 2\lambda)\h
\ \ \ , 
\label{eq:hamiltonian} \\
{\cal H}^a &=& -2D_b \pi^{ab} \ \ \ . 
\label{eq:momentum}
\end{eqnarray}
Here, $\lambda$ is the cosmological constant which is set to be
 zero if it is not being considered.

Taking   ${\cal H} ={\cal H} (\h) $, a canonical 
pair $(\sqrt{h}, \pi / \h)$ can be chosen to be gauge-fixed.  
One natural way to fix the gauge is to impose a 1-parameter 
family of gauge-fixing conditions, 
\begin{equation}
\chi_{{}_1} :=   {\pi \over \h} - \s =0\ \ \ (\exists \s \in 
{\bf R})\ \ ,
\label{eq:gauge1}
\end{equation}   
where $\s$ is a spatially constant parameter 
(York's gauge~\cite{YORK}).
 Let us  make clear the meaning of the gauge Eq.(\ref{eq:gauge1}).
 
 We adopt the following 
notations; $(P_1^\dagger w)^a := -2D_b w^{ab}$ for a symmetric 
 traceless tensor $w^{ab}$;
 $\tilde{\pi}^{ab}:= \pi^{ab}-{1\over 2}\pi h^{ab}$ is the   
traceless part of $\pi^{ab}$ and in particular 
$\tilde{\pi}^{'ab}$ stands  for
$\tilde{\pi}^{'ab} \notin \ker P_1^\dagger$.

Now, let $Q:= {\pi \over \h}$ and 
$Q':= \int_{\Sigma} d^2 x\ \h \ Q \big/ 
                    \int_{\Sigma} d^2 x \h $, 
which is the spatially constant component of $Q$. 
Therefore, 
${\cal P}' \ (\cdot)
:= \int_{\Sigma} d^2 x\ \h \  (\cdot) 
      \big/ \int_{\Sigma} d^2 x \h $
 forms a linear map which projects $Q$ to $Q'$. 
 On the other hand, 
 ${\cal P}= 1- {\cal P}'$ projects $Q$ to 
 its spatially varying component. 
 Note that $({\cal P}Q, {\cal P}' Q)=0$ w.r.t. the 
 natural inner product ({\it Appendix} $A$). 
 Then, Eq.(\ref{eq:gauge1}) 
  can be recast as
\begin{equation}
\chi_{{}_1} := {\cal P} \left( {\pi \over \h} \right)=0\ \ \ .
\label{eq:gauge1'}
\end{equation}

 We note that ${\cal H}^a = -2D_b \tilde{\pi}^{ab} 
=: (P_1^\dagger \tilde{\pi})^{ab}$ 
under the condition of Eq.(\ref{eq:gauge1'}).
Taking  
${\cal H}^a ={\cal H}^a (\tilde{\pi}^{'ab})$,  a  pair 
$(h_{ab}/\h, \tilde{\pi}^{'ab}\h)$ shall  be gauge-fixed.      
Thus  we choose as a gauge-fixing condition, 
\begin{equation}
\chi_{{}_2} := {h_{ab} \over \h}-  \hh (\tau^A) =0\ \ \ 
(\exists \tau^A \in {\cal M}_g) \ \ \ ,
\label{eq:gauge2}
\end{equation}
where $\hh$ is a $m$-parameter family of reference metrics 
($m=2, 6g-6$ for $g=1$, $g \geq 2$, respectively) s.t. 
$\det \hh =1 $; $\tau^A$ ($A=1, \cdots, m$) denote the 
Teichm\"uller parameters parameterizing the moduli space 
${\cal M}_g$ of $\Sigma$~\cite{HAT}.  
 
At this stage, we  recall~\cite{HAT} that 
 a general  variation of $h_{ab}$ can be decomposed as 
$\d h_{ab}= \d_W h_{ab} + \d_D h_{ab} +\d_M h_{ab}$, where 
$\d_W h_{ab}$ is the trace part of $\d h_{ab}$ 
(Weyl deformation), 
$\d_D h_{ab}=(P_1v)_{ab}:=D_a v_b +  D_b v_a - D_c v^c h_{ab}$ 
for ${\exists v^a}$ (the traceless part of a diffeomorphism), 
and 
 $\d_M h_{ab}=
 {\cal T}_{Aab} \d \tau^A := 
 ({\partial h_{ab} \over \partial \tau^A} 
  -{1\over 2}h^{cd}{\partial h_{cd} \over 
               \partial \tau^A }h_{ab}) \d \tau^A$  
  (the traceless part of 
 a moduli deformation).\footnote{
      Needless to say, these quantities are defined for $h_{ab}$,
      a spatial metric induced on $\Sigma$. Therefore, 
      under the condition  
      (\ref{eq:gauge2}), they are calculated  using 
      $\h \hh (\tau^A)$, and 
      not just $\hh (\tau^A)$.
      \label{footnote:remark}} 
  It is easy to show 
   that~\cite{HAT}, the adjoint of $P_1$ w.r.t. the natural 
   inner product 
({\it Appendix} $A$) becomes  $(P_1^\dagger w)^a :=-2D_b w^{ab}$,
 acting on  a symmetric  traceless 
 tensor $w^{ab}$. [Therefore the notation 
  ``$P_1$" is compatible with the  
  notation ``$P_1^\dagger $" introduced just after 
  Eq.(\ref{eq:gauge1}).]

Now, the meaning of the gauge Eq.(\ref{eq:gauge2}) is 
as follows. 
The variation of $h_{ab}/\h$ in the neighborhood of
 $\hh (\tau^A)$ is expressed as\footnote{
     The symbol $\d \left\{ \cdot \right\}$ shall
     be used to represent a 
    variation whenever there is a possibility  
    of  being confused  
     with the delta function $\d (\cdot)$.
    } 
\[
\d \left\{ h_{ab}/\h \right\} 
= \d_D \left\{ h_{ab}/\h \right\} + \d_M \left\{ h_{ab}/\h
    \right\}\ \ \ .
\]
 Let  $Riem_1 (\Sigma)$ denote the  space of unimodular 
 Riemannian metrics on $\Sigma$. 
 We introduce projections defined on  the tangent space of 
 $Riem_1 (\Sigma)$ at $\hh (\tau^A)$,\newline
 $T_{\hh (\tau^A)} (Riem_1 (\Sigma))$: 
\begin{eqnarray*}
  {\cal P}_D \left( \d \left\{ h_{ab}/\h \right\} \right)  
   &=& \d_D \left\{ h_{ab}/\h \right\} \ \ \ , \\
  {\cal P}_M \left( \d \left\{ h_{ab}/\h \right\} \right) 
   &=& \d_M \left\{ h_{ab}/\h \right\} \ \ \ , \\
    {\cal P}_D + {\cal P}_M &=&1 \ \ \ .
\end{eqnarray*}
Then, the gauge Eq.(\ref{eq:gauge2}) is recast as
\begin{equation}
\chi_{{}_2} = {\cal P}_D \left( \d \left\{ h_{ab}/\h \right\} 
           \right) =0\ \ \ .
\label{eq:gauge2'}
\end{equation}        
On $Riem_1 (\Sigma)$ we can introduce a system of coordinates 
in the neighborhood of each $\hh (\tau^A)$. Then 
$[d h_{ab}]$ in  Eq.(\ref{eq:partition}) is expressed as 
$[d\h][d \d\left\{ h_{ab}/\h \right\}]$. 
[It is easy to show that 
the Jacobian factor associated with this change of variables 
is unity.]

Finally let us discuss about the integral domain for $N_a$ in 
Eq.(\ref{eq:partition}) for the case of $g=1$.\footnote{
                The author thanks S. Carlip for valuable remarks 
                on this point.}
Let us note that, under the gauge Eq.(\ref{eq:gauge1'}), we get 
\[
\int_{\Sigma} d^2 x \ N_a {\cal H}^a = 
2 \int_{\Sigma} d^2 x (P_1 N)_{ab} \tilde{\pi}^{ab}\ \ \ .
\]
Thus, when $N_a \in \ker P_1$,  
$N_a$ does not work as a Lagrange multiplier 
enforcing the momentum constraint 
Eq.(\ref{eq:momentum}). 
Then there are two possible options for 
the path-integral domain of 
$N_a$:
\def\labelenumi{(\theenumi)}
\def\theenumi{\alph{enumi}}
\begin{enumerate}
\item All of the vector fields on $\Sigma$, including $\ker P_1$. 
     \label{item:a}
\item All of the vector fields on $\Sigma$, except for  $\ker P_1$.
     \label{item:b}
\end{enumerate}

If we choose the option (\ref{item:a}), the integral over 
$N_a$ in Eq.(\ref{eq:partition}) yields the factors 
$\det{}^{1/2} (\varphi_\a , \varphi_\b)
\d (P_1^\dagger \tilde{\pi})$.  
Here $\{ \varphi _\a \}_{\a =1,2}$ is a basis of  $\ker P_1$ 
 for the case of $g=1$.\footnote{
           Let us recall that $\dim \ker P_1 = 6, 2$ 
           and $0$ for $g=0$, $g=1$ and 
           $g \geq 2$, respectively. On the other hand, 
           $\dim \ker P_1^\dagger = 0, 2$ 
           and $6g-6$ for  $g=0$, $g=1$ and 
           $g \geq 2$, respectively. 
           There is a relation 
           $\dim \ker P_1 - \dim \ker P_1^\dagger =6-6g$ 
           (Riemann-Roch Theorem)~\cite{HAT}. 
            [Throughout this paper, $\dim W$ indicates 
             the real dimension of a  space $W$, regarded as 
             a  vector space over $\bf R$.]}
 The factor $\det{}^{1/2} (\varphi_\a , \varphi_\b)$
  appears here since it is proportional to the volume of 
  $\ker P_1$ w.r.t. the natural inner product. 
  
  If we choose the option (\ref{item:b}), the integral 
  over $N_a$ yields just a factor 
  $\d (P_1^\dagger \tilde{\pi})$.

Integrating over the Lagrange multipliers $N$ and $N_a$, 
(\ref{eq:partition}) reduces to 
\begin{equation}
 Z={\cal N} \int [dh_{ab}][d\pi^{ab}]\  {\cal B}\  
 \d ({\cal H}) \d({\cal H}^a)
   \exp i\int dt \int_{\Sigma} d^2x \  \pi^{ab} \dot{h}_{ab}
   \ \ \ ,
  \label{eq:partition2}
  \end{equation}
where
\[
{\cal B} = \cases{
           \det{}^{1/2} (\varphi_\a , \varphi_\b)& 
              {\rm when}\  g=1\  
              {\rm with\  the\  option\  (\ref{item:a})}         
                                                        \cr
          1  & otherwise      \cr
                }\ \ \ .
\]

 According to the Faddeev-Popov procedure~\cite{FAD}, we insert 
into the right-hand side of Eq.(\ref{eq:partition2}) the factors
\[
 |\det \{{\cal H}, \chi_{{}_1} \}| 
 |\det \{{\cal H}^a, \chi_{{}_2} \}|
 \d(\chi_{{}_1}) \d(\chi_{{}_2}) \ \ \ .
\]
Note that, because 
$\{ \int v_a {\cal H}^a, \chi_{{}_1} \} = -v_a {\cal H}^a -v^c 
  D_c \chi_{{}_1}=0$ 
  {\it mod} 
  ${\cal H}^a=0$ and $\chi_{{}_1} =0$,  
the Faddeev-Popov determinant separates into 
two factors as above.\footnote{
      For notational neatness, the symbol of absolute value 
      associated with the Faddeev-Popov 
      determinants shall be omitted for most 
      of the cases. } 
The determinants turn to simpler expressions if we note the 
canonical structure of our system;
\begin{eqnarray*}
\int_{\Sigma} d^2x \ \pi^{ab} \d h_{ab} 
&=& \int_{\Sigma} d^2x \ 
  \left( \tilde{\pi}^{ab} + {1\over 2} \pi  h^{ab} \right)
  \left( \d_W h_{ab} + \d_D h_{ab} +\d_M h_{ab} \right) \\
 &=& \int_{\Sigma} d^2x \ 
    \left( { \pi\over \h}  \d \h + 
    (P_1^\dagger \tilde{\pi}')^a v_a 
      + \tilde{\pi}^{ab} \d_M h_{ab} \right)\ \ \ .
\end{eqnarray*} 
Thus, 
\begin{eqnarray*}
\det \{{\cal H}, \chi_{{}_1} \} &=& {\partial {\cal H} \over 
\partial \h }\ \ \ ,  \\
\det  \{{\cal H}^a, \chi_{{}_2} \} &=& 
\left( \det{\partial {\cal H}^a \over \partial 
\tilde{\pi}^{'ab}}  
\right) \cdot {\partial \chi_{{}_2}  \over  
     \partial \left( \d_D h_{ab} \right)} 
     = \det{}' P_1^\dagger \  \h
     \ \ \ .
\end{eqnarray*} 
Thus we get 
\begin{eqnarray}
Z &{}&={\cal N}  
\int [ d \h \ \ d \left( \d \left\{ h_{ab}/\h \right\} \right)\ 
         d\left( \pi/\h \right) \   d \tilde{\pi}^{ab}]\ {\cal B}  
         \nonumber \\ 
  &{}&       \ {\partial {\cal H} \over \partial \h }\   
   \det{}' P_1^\dagger \  \h \ 
    \d ({\cal H})\  \d (P_1^\dagger \tilde{\pi})\  
     \d \left( {\cal P} \left( \pi / \h \right) \right) \ 
     \d \left( {\cal P}_D \left( \d \left\{ h_{ab}/\h \right\} 
     \right) 
         \right) \nonumber \\
  &{}& \exp i \int dt \int_{\Sigma} d^2 x \ 
   \big( \tilde{\pi}^{ab} + {1\over 2} \pi  h^{ab} \big)
   \dot{h}_{ab}    \ \ \ .
  \label{eq:partition3}
 \end{eqnarray} 

We can simplify the above expression. 
First of all, the path integral w.r.t. 
$ \pi / \h $ in Eq.(\ref{eq:partition3})
 is of the form 
 \[
 I_1 = \int d \left( \pi / \h  \right) 
 \d \left( {\cal P} \left( \pi / \h \right) \right) 
 F\left( \pi / \h  \right)\ \ \ ,
 \]
 so that Eq.(\ref{eq:keyformula}) in $\it Appendix$ $B$ 
  can be applied. Note that $\ker {\cal P}=$a 
   space of spatially constant
  functions, which forms a 1-dimensional vector space over 
  $\bf R$. Now  $\dim \ker {\cal P}=1$, so that   $dp_A$ and 
  $d (p_A \vec{\Psi}^A)$ are equivalent, 
  following the notation in 
  $\it Appendix$ $B$.  Furthermore 
   $\cal P$ is a projection. Thus no extra Jacobian factor 
  appears in this case. 
  Thus we get 
  \[ I_1 = \int [d\s] F({\cal P} \left( \pi / \h \right)=0, \s )
  \ \ \ ,
  \] 
  where $\s$ denotes  a real parameter parameterizing 
  $\ker {\cal P}$. 
  
  Second, the path integral w.r.t. 
  $ \d \left\{ h_{ab}/\h \right\} $ 
  is of the form 
\[ I_2=\int d \d \left\{ h_{ab}/\h \right\} 
  \d \left( {\cal P}_D \left( \d \left\{ h_{ab}/\h \right\} 
    \right) 
   \right) G\left( \d \left\{ h_{ab}/\h \right\} \right) \ \ \ .
\]
   Note that $\ker {\cal P}_D = \d_M \left\{ h_{ab}/\h \right\}
    = \sqrt{h^{-1}} \d_M h_{ab}$ and $\det{}' {\cal P}_D =1$. 
    Let 
    $\{ \xi ^A \}$ $(A=1, \cdots , \dim \ker P_1^\dagger  )$ 
    be a basis of $\ker {\cal P}_D $. Then the factor 
    $\det{}^{1/2} (\xi^A , \xi^B)$ 
    (see Eq.(\ref{eq:keyformula})) 
    is given as 
  \begin{equation}
  \det{}^{1/2} (\xi^A , \xi^B)=
  \det ({\cal T}_A , \Psi^B) \det{}^{-1/2} (\Psi^A, \Psi^B) 
 \sqrt{h^{-1}}\ \ ,
  \label{eq:detxi}
  \end{equation}
 where $\{ \Psi^A \}$ $( A=1, \cdots , \dim \ker P_1^\dagger )$ 
 is a basis of $\ker P_1^\dagger$. 
 This expression results in as follows. 
 Carrying out  a standard manipulation~\cite{HAT,CAR2,MS1},
 \footnote{
     Because $P_1^\dagger$ is a Fredholm operator on  a space of 
     symmetric traceless tensors ${\cal W}$, ${\cal W}$ can be
      decomposed as 
     ${\cal W}= Im P_1 \oplus \ker P_1^\dagger$~\cite{NAKA}. 
     Therefore  ${\cal T}_{Aab} \d \tau^A \in {\cal W}$ is 
     uniquely decomposed in the form of 
     $P_1 u_0 + ({\cal T}_A , \Psi^B) 
     {(\Psi^{\cdot}, \Psi^{\cdot})^{-1}}_{BC} 
     \Psi^C \d \tau^A$. 
     Then, $(P_1 \tilde{v})_{ab}:= (P_1 (v+u_0))_{ab} $.}
 \begin{eqnarray}
 \d h_{ab}&=& {\d \h \over  \h} h_{ab} + (P_1 v)_{ab} 
                 + {\cal T}_{Aab} \d \tau^A \nonumber \\
       &=&  {\d \h \over  \h} h_{ab} + (P_1 \tilde{v})_{ab}
  + ({\cal T}_A , \Psi^B) 
  {(\Psi^{\cdot}, \Psi^{\cdot})^{-1}}_{BC} 
  \Psi^C \d \tau^A\ \ \ . 
\label{eq:delh}
 \end{eqnarray}
For the present purpose, the first and second terms are 
set to be zero. [See the 
footnote \ref{footnote:remark}.] 
According to {\it Appendix} $A$, then, 
it is easy to get Eq.(\ref{eq:detxi}).  Then with the help of 
Eq.(\ref{eq:keyformula}), we get 
\[
I_2 = \int d\tau^A\ 
\det ({\cal T}_A , \Psi^B) \det{}^{-1/2} (\Psi^A, \Psi^B) 
 \sqrt{h^{-1}}\ 
 G \left( \d_D \left\{ h_{ab}/\h \right\}=0, \tau^A \right)
  \ \ \ .
\]
Here we understand that the integral domain for 
$\int d\tau^A$ is on  the moduli space  
${\cal M}_g$,  and not the Teichm\"uller space, which 
is the universal covering space of ${\cal M}_g$~\cite{HAT}.
 This is clear 
because $\tau^A$ appears in the integrand $G$ only through 
$\hh$ (Eq.(\ref{eq:gauge2})).  

 We note that 
 the kinetic term in Eq.(\ref{eq:partition3}) becomes 
\begin{eqnarray*}
  \int_{\Sigma} d^2 x \ 
  { (\tilde{\pi}^{ab}+{1\over 2}\pi  h^{ab})
    \dot{h}_{ab}  }_{|_{ \chi_{{}_1}=\chi_{{}_2}=0 } } 
   &=& \int_{\Sigma} d^2 x \  
  \left( \tilde{\pi}^{ab} + {1\over 2} \s \h h^{ab} \right) 
  {\dot{h}_{ab}}
  |_{ \chi_{{}_2} =0 }  \\
&{}&  =  
\left( \tilde{\pi}^{'ab} + p_A \Psi^{Aab} \h, {\cal T}_{Bcd} 
\right) \dot{\tau}^B +\s \dot{V}  \ \ \ .
\end{eqnarray*}
Here  $V:=\int_{\Sigma} d^2 x\ \h$, which is interpreted as a 
2-volume (area) of $\Sigma$.  
[See {\it Appendix} $A$ for the inner product of densitized 
quantities.] 

Finally, the path integral w.r.t. $ \tilde{\pi}^{ab}$ in 
Eq.(\ref{eq:partition}) 
is of the form 
\[
I_3 = \int d \tilde{\pi}^{ab}\ \d (P_1^\dagger \tilde{\pi})\ 
        H \left( \tilde{\pi}^{ab} \right)\ \ \ .
\]
Using Eq.(\ref{eq:keyformula}),  this is recast as
\[
I_3=\int dp_A\ \det{}^{1/2} (\Psi^A, \Psi^B)  
(\det{}' P_1^\dagger)^{-1}
 H \left( \tilde{\pi}^{'ab}=0, p_A  \right) \ \ \ .
\]

Combining the above  results for $I_1$, $I_2$ and $I_3$, 
the expression in Eq.(\ref{eq:partition3}) is recast as
\begin{eqnarray}
Z=&{}& {\cal N}  \int [d\h \ d \s \ d\tau^A dp_A ] \   
 {\partial {\cal H} \over \partial \h }\  \d ({\cal H})\  
{ \det ({\cal T}_A , \Psi^B) \over  \det{}^{1/2} 
(\varphi_\a , \varphi_\b)}\ {\cal B}
  \nonumber \\
&{}& \exp i \int_{\Sigma}dt\ 
\{ p_A (\Psi^A, {\cal T}_B)\dot{\tau}^B 
+ \s \dot{V} \} \ \ \ .
\label{eq:partition4}
\end{eqnarray}
  The reason why the factor 
 $\det{}^{-1/2} (\varphi_\a , \varphi_\b)$ appears in 
 Eq.(\ref{eq:partition4}) for $g=1$ shall be discussed  
 below. [For the case of $g \geq 2$,  
the factor $\det{}^{-1/2} (\varphi_\a , \varphi_\b)$  
should be set to unity.] 
Without loss of generality, we can choose a basis of 
$\ker P_1^\dagger$, 
 $\{ \Psi^A \}$, as to satisfy  
 $({\cal T}_A , \Psi^B)={\d _A}^B $.

Under our gauge choice, the equation 
${\cal H}=0$ considered  as being an equation for $\h$, 
 has an unique solution, 
$\h = \h (\cdot\ ; \s, \tau^A, p_A)$, for fixed 
$\s$, $\tau^A$, and $p_A$~\cite{MONC}.
We therefore obtain
\begin{equation}
Z= {\cal N}  \int [d\s  \ d\tau^A dp_A ] \   
 \det{}^{-1/2} (\varphi_\a , \varphi_\b) \ {\cal B}
 \  \exp i \int\ dt\ ( p_A \dot{\tau}^A 
+ \s \dot{V}(\s, \tau^A, p_A) ) \ \ \ ,
\label{eq:partition5}
\end{equation}
where  
$V(\s, \tau^A, p_A):=\int_{\Sigma} d^2 x\ \h(x; \s, \tau^A, p_A)$,
which  is regarded as a function of 
$\s$, $\tau^A$, and $p_A$.

It is clear that there is still the invariance under the 
reparameterization $t \rightarrow f(t) $ remaining in 
Eq.(\ref{eq:partition5}).     From the geometrical viewpoint, 
this corresponds to the 
freedom in the way of  labeling the  time-slices 
defined by Eq.(\ref{eq:gauge1}).
 [This point is also clear in the analysis of Ref.\cite{MONC}. 
 The treatment of this 
  point seems somewhat obscure in the analysis of 
  Ref.\cite{CAR2}.] 
The present system illustrates that the time reparameterization 
invariance still remains even after choosing the time-slices 
(Eq.(\ref{eq:gauge1}) or Eq.(\ref{eq:gauge1'})). 

Eq.(\ref{eq:partition5}) is equivalent to 
\begin{eqnarray}
Z &=& {\cal N}  \int [d\s  \ dp_{\s} d\tau^A dp_A ] [dN'] \   
 \det{}^{-1/2} (\varphi_\a , \varphi_\b) \ {\cal B} \nonumber \\
&{}& \  \exp i \int \ dt\ \{ p_A \dot{\tau}^A + p_{\s} \dot{\s} 
   - N'(p_{\s} + V(\s, \tau^A, p_A))  \} \ \ \ ,
\label{eq:partition6}
\end{eqnarray}
where the integration by parts is understood. 
This system has a similar structure to a system of 
a relativistic 
particle and a system of a non-relativistic particle in a 
parameterized form~\cite{HART}. We shall discuss this 
point in detail in the final section. 
One can  gauge-fix the reparameterization symmetry by choosing 
$\s=t$, i.e. by imposing a condition $\chi = \s -t =0$.
The Faddeev-Popov procedure~\cite{FAD} in this case 
reduces to simply inserting $\d (\s-t)$ into
 Eq.(\ref{eq:partition6}). 
Thus we obtain 
\begin{equation}
Z={\cal N} 
 \int [d\tau^A dp_A] \  {\cal A} \ 
\exp i \int d\s ( p_A\  d{\tau}^A / d\s- V(\s, \tau^A, p_A))
\ \ \ .
\label{eq:partition7}
\end{equation}
 Here 
\[
{\cal A} = \cases{
           \det{}^{-1/2} (\varphi_\a , \varphi_\b)& 
              {\rm when}\  g=1\  
              {\rm with\  the\  option\  (\ref{item:b})}         
                                                        \cr
          1  & otherwise      \cr
                }\ \ \ .
\]
 
 Looking at the exponent  in Eq.(\ref{eq:partition7}), 
 we see  that    
 $V(\s, \tau^A, p_A)$ plays the role of 
a time-dependent Hamiltonian in the present gauge~\cite{MONC}.
We see that the partition function formally defined by  
Eq.(\ref{eq:partition})  is equivalent to  the 
partition function defined by  taking  
the reduced system as a starting point, as can be read off in 
Eq.(\ref{eq:partition7}).  However, there is one point to 
be noted. 
For the case of $g=1$ with the option (\ref{item:b}), 
the factor $\det{}^{-1/2} (\varphi_\a , \varphi_\b)$ appears. 
This factor 
can cause a nontrivial effect. We shall come back to this point
 in the next section. 
Typically, this factor can be a function of $V(\s, \tau^A, p_A)$
 (see below, Eq.(\ref{eq:factor1})).
On the contrary, for the case of $g = 1$ with the 
option (\ref{item:a}) and for the case of $g \geq 2$,   this factor 
does not appear.  We especially note that, for 
the case of $g = 1$ with the option (\ref{item:a}), 
the factor $\det{}^{1/2} (\varphi_\a , \varphi_\b)$ 
coming from $\int [d N_a]$ cancels with the same factor 
appeared in Eq.(\ref{eq:partition4}).\footnote{
                 The author thanks S. Carlip for very helpful 
                 comments on this point.}

Let us discuss the factor 
$\det{}^{-1/2} (\varphi_\a , \varphi_\b)$
 in Eq.(\ref{eq:partition7}).
 
 In the case of $g=1$, the space $\ker P_1$, which is 
 equivalent to a space of 
 conformal Killing vectors, is nontrivial. Now  a special class
  of Weyl deformations represented as 
 $\d_W h_{ab}= D \cdot v_0\  h_{ab}$, where $v_0 \in \ker P_1$, 
 is  translated into a diffeomorphism: $D \cdot v_0\ h_{ab}
 = (P_1 v_0)_{ab} + D \cdot v_0\ h_{ab} 
 = {\cal L}_{v_0} h_{ab}$. 
  [Here ${\cal L}_{v_0}$ denotes the Lie derivative w.r.t. $v_0$.]
   Thus, $\d_W h_{ab}= D \cdot v_0\ h_{ab}$, $v_0 \in \ker P_1$ 
   is  generated by ${\cal H}^a$ along the gauge orbit.  
 Therefore it  should be removed 
 from the integral domain  for $\int [d\h]$ in 
 Eq.(\ref{eq:partition3}).
  One easily sees that the volume of $\ker P_1$, which should be 
  factorized out from the whole volume of 
   the  Weyl transformations,
   is proportional to 
  $\det{}^{1/2} (\varphi_\a , \varphi_\b)$. Therefore the 
  factor  $\det{}^{-1/2} (\varphi_\a , \varphi_\b)$ appears in 
Eq.(\ref{eq:partition7}).

There is another way of  explaining    the factor 
$\det{}^{-1/2} (\varphi_\a , \varphi_\b)$~\cite{MS1}. Let us 
concentrate on the diffeomorphism 
invariance in Eq.(\ref{eq:partition2})  characterized by 
${\cal H}^a=0$.
The Faddeev-Popov determinant associated with this  
invariance  can be related to the Jacobian 
for   the change $h_{ab} \rightarrow (\h, v^a, \tau^A)$. 
By the same kind of argument as in  Eq.(\ref{eq:delh}),
one finds the Faddeev-Popov determinant to be 
\[
\Delta_{FP} 
= \det ({\cal T}_A , \Psi^B)\  \det{} ^{-1/2} (\Psi^A, \Psi^B)
 (\det{}' P_1^\dagger P_1)^{1/2}\ \ \ .
\]
One way of carrying out  the  Faddeev-Popov procedure is to 
insert \\ 
$1=\int d\Lambda \det{ \partial \chi \over \partial \Lambda} 
\d (\chi)$ 
into the path-integral formula in question, where $\chi$ is 
a gauge-fixing function and $\Lambda$ is a gauge parameter.  
Then   the  path integral  in Eq.(\ref{eq:partition2}) reduces 
to  the form
\begin{eqnarray*}
I&=&\int [dh_{ab} ] [d\h\ dv^a\ d\tau^A][d\ast]\  
\d (h_{ab}-\h \hh) 
\ f(h_{ab}) \\
&=& \int  [d\h\ dv^a\ d\tau^A][d\ast]\  f(\h \hh)\ ,
\end{eqnarray*}
where $[d\ast]$ stands for all of the remaining integral 
measures including $\Delta_{FP}$.

Now, we need to   factorize out $V_{Diff_0}$, 
the whole volume of diffeomorphism homotopic to $1$. This 
volume is related to $\int [dv^a]$ as 
$V_{Diff_0} = (\int [dv^a])\cdot V_{\ker P_1}$, where 
${V_{\ker P_1}} \propto 
 \det{}^{1/2} (\varphi_\a , \varphi_\b)$~\cite{HAT}. 
 Here we note that 
 $\ker P_1$ is not included in the  integral domain of 
 $\int [dv^a]$:
the diffeomorphism associated with $\forall v_0 \in \ker P_1$ 
is  translated into a Weyl transformation, as
 ${{\cal L}_{v_0}h}_{ab}=(P_1 v_0)_{ab} + D \cdot v_0\ h_{ab}
 =D \cdot v_0\ h_{ab} $ [it is noteworthy that  this  argument
  is  reciprocal to the  previous one], 
 so that  it is already counted in $\int [d\h]$.
 In this manner we get 
 \[
 I= V_{Diff_0} \int { [d\h\  d\tau^A] [d\ast] \over 
  \det{}^{1/2} (\varphi_\a , \varphi_\b)}\ f(\h \hh)\ \ \ .   
 \]
 In effect, the volume of $\ker P_1$ has been removed 
 from the whole volume of the Weyl transformations, which is 
 the same result as the one in the previous argument. 
 [Again, for the case of  $g \geq 2$,  
the factor $\det{}^{-1/2} (\varphi_\a , \varphi_\b)$  
should be set to unity.] 
Furthermore, by factorizing the entire volume of
 diffeomorphisms, $V_{Diff}$, and not just $V_{Diff_0}$, 
 the integral domain for 
$\int [d\tau^A]$ is reduced to the moduli space, 
${\cal M}_g$~\cite{HAT,MS1}. The intermediate step of 
factorizing $V_{Diff_0}$ is necessary since the $v^a$'s are 
labels parameterizing  the  tangent 
space of $Riem(\Sigma)$, the  space of 
all Riemannian metrics on $\Sigma$.


\section{Analysis of the  $g=1$ case  }
\label{section:III}

We now investigate  
 how the reduced canonical system in the 
 $(\tau, V)$-form~\cite{HOS} 
  comes out in the partition function when $g=1$.

 To begin with, let us recover $\int [d\h]$ and $\int [d\s]$ 
 in Eq.(\ref{eq:partition7}), yielding
\begin{eqnarray}
Z &=& {\cal N} 
  \int [d\h][d\s][d\tau^A dp_A] \ {\cal A} \ 
          {\partial {\cal H} \over \partial \h }\
    \d ({\cal H}) \  \d (\s-t) \nonumber \\
&{}& \ \exp i \int dt 
 \{ p_A\  {d {\tau}^A \over dt}  + \s {d \over dt} 
 V(\s, \tau^A, p_A) \}
                             \ \ \ .
\label{eq:partition7b}
\end{eqnarray} 
Eq.(\ref{eq:partition7b}) is  of  the form 
\begin{equation}
  I=\int [d\h][d\ast]\ 
   {\partial {\cal H} \over \partial \h }\
    \d ({\cal H})\ f(\h)\ \ \ ,
\label{eq:I}
\end{equation}
  where $[d\ast]$  
  stands for all of the remaining integral measures. 

Now it is shown that for $g=1$ the simultaneous differential
 equations Eq.(\ref{eq:hamiltonian}), Eq.(\ref{eq:momentum}), 
 Eq.(\ref{eq:gauge1}) (or  Eq.(\ref{eq:gauge1'}))
  and Eq.(\ref{eq:gauge2}) (or Eq.(\ref{eq:gauge2'})) 
  have a unique solution 
 for $\h$, which is spatially constant, 
 $\h _0
 :=F(\tau^A, p_A, \s)$~\cite{MONC}.   
 Thus the integral 
  region for $\int [d\h]$ in Eq.(\ref{eq:I}) can be restricted 
  to ${\cal D}=\{\h| \h ={\rm spatially \  constant} \}$. 
  Let us note that 
  $\h$ is the only quantity that in principle 
  can depend on spatial coordinates in
   Eq.(\ref{eq:partition7b}).  
  Accordingly, only the spatially constant components of the 
  arguments of the integrand contribute to the path integral. 
  
  Thus,
  \begin{eqnarray*}
  I&=&\int [d\ast]\  f(\h _0)  \\
   &=& \int_{\cal D} [d\h][d\ast]\  
   \left\{ \int_{\Sigma} d^2x\ \h \   
    {\partial {\cal H} \over \partial \h }\ 
                  {\Bigg/} \int_{\Sigma}d^2 x\  \h \right\} 
                  \times \\ 
  &{}&\times 
  \d \left(\int_{\Sigma} d^2x\ {\cal H}\ {\Bigg/} 
   \int_{\Sigma} d^2x \right) \ f(\h) \\
   &=& \int _{\cal D} ([d\h] \int_{\Sigma} d^2x)[d\ast]\ 
           {\partial  H \over \partial V }\  
           \d (H) \  \tilde{f}(V) \\
   &=& \int [dV][d\ast]\  
           {\partial  H \over \partial V }\  
           \d (H)\  \tilde{f}(V) \\
   &=& \int [dV][d\ast][dN']\  {\partial  H \over \partial V }\ 
        \tilde{f}(V)\ \exp-i\int dt\ N'(t) H(t)\ \ \ ,
   \end{eqnarray*}
   where $H:=\int_{\Sigma} d^2x\ {\cal H}$, 
    $V:=\int_{\Sigma} d^2x\ \h$  and $\tilde{f}(V):= f(\h)$. 
    The prime symbol in $N'(t)$ is to emphasize that 
    it is spatially constant. 
 
  Thus we see that Eq.(\ref{eq:partition7b}) is equivalent to 
\begin{eqnarray}
Z&=&{\cal N} 
 \int   [dV\ d\s][d\tau^A\ dp_A] [dN'] \  {\cal A} \ 
  {\partial  H \over \partial V }  \ \d (\s-t) \nonumber \\
&{}& \exp i \int dt\  ( p_A\  \dot{\tau}^A + \s \dot{V}- N'H)
  \ \ \ ,
  \label{eq:partition8}
  \end{eqnarray}
  where $V$ and $N'$ are spatially constant, and 
   $H$ is the  reduced Hamiltonian in the $(\tau, V)$-form.  
   [See below, Eq.(\ref{eq:reducedaction}).]

We  choose as a gauge condition 
(see Eq.(\ref{eq:gauge2}))~\cite{MS1},
\[
h_{ab}=V \hh \ , \ \ \ 
\hh  ={ 1\over  \tau^2 }
\pmatrix{ 1     &  \tau^1  \cr
         \tau^1 &  |\tau|^2 \cr}\ \ \ ,
 \]
 where $\tau:= \tau^1 + i \tau^2$ 
 and $\tau^2 >0$.\footnote{
 Throughout this section, $\tau^2$ always indicates the second 
 component of $(\tau^1, \tau^2)$, and never the square of 
 $\tau:= \tau^1 + i \tau^2$.
 }
 Here we have already replaced $\h$ with  $V=\int_{\Sigma} \h$, 
 noting that $\h$ is spatially constant for the case of $g=1$.
  Then, it is straightforward to get 
\[
{\cal T}_{1ab}={V \over  \tau^2}
   \pmatrix { 0      &  1       \cr
                            1      &  2\tau^1      \cr},\ \ 
{\cal T}_{2ab}={V \over  (\tau^2)^2}
   \pmatrix { -1           &  -\tau^1                    \cr
              -\tau^1      &  (\tau^2)^2-(\tau^1)^2      \cr}\ \ .
\]
 [See the paragraph next to the one including 
 Eq.(\ref{eq:gauge2})  for the definition of $\{ {\cal T}_A \}$.]

As a basis of $\ker P^\dagger_1$, $\{\Psi^A \}_{A =1,2}$, the 
fact that 
$P^\dagger_1({\cal T}_A)_a:=-2D_b{{\cal T}_{Aa}}^b
 =-2\partial_b {{\cal T}_{Aa}}^b=0$ simplifies the situation.
  We can choose as  $\{\Psi^A \}_{A =1,2}$  
\[
\Psi^1_{ab}={1 \over 2 }
   \pmatrix { 0      &  \tau^2       \cr
              \tau^2      &  2\tau^1 \tau^2     \cr},\ \ 
\Psi^2_{ab}={ 1 \over 2 }
   \pmatrix { -1           &  -\tau^1                    \cr
              -\tau^1      &  (\tau^2)^2-(\tau^1)^2      \cr}\ \ ,
\]
which satisfy $( \Psi ^A, {\cal T}_B) ={\d^A}_B$.

Now let us consider in detail the case of the option 
(\ref{item:b}) (\S \ref{section:II}). 
In this case the factor $\cal A$
becomes ${\cal A} = \det{}^{-1/2} (\varphi_\a , \varphi_\b)$. 
As a basis of $\ker P_1$, $\{\varphi_\a \}_{\a =1,2}$, 
we can take spatially constant vectors  because 
$D_a=\partial_a$ for the  metric in question, and  
because constant vectors are compatible with the condition for 
the allowed vector fields on $T^2$. 
 [Note the fact that the Euler characteristic of $T^2$ 
 vanishes,  along with the Poincar\'e-Hopf 
 theorem~\cite{DUB}.]   
Therefore, let us take as 
\[
{\varphi_1}^a = \lambda_1 \pmatrix{1 \cr
                      0 \cr}\ , \ \ 
{\varphi_2}^a = \lambda_2 \pmatrix{0 \cr
                      1 \cr}\ , \ \ 
 \] 
where $\lambda_1$ and $\lambda_2$ are spatially constant 
factors. Then, we get 
\[
 (\varphi_\a, \varphi_\b)=
\pmatrix{ {\lambda_1}^2 V^2 /\tau^2 &
                   \lambda_1 \lambda_2 V^2 \tau^1 /\tau^2 \cr
      \lambda_1 \lambda_2 V^2 \tau^1 /\tau^2 & 
             {\lambda_2}^2 V^2 |\tau|^2/\tau^2 \cr
        }\ \ \ .
\]
Thus, we obtain
\begin{equation} 
\det{}^{1/2} (\varphi_\a, \varphi_\b) = |\lambda_1 \lambda_2|
 V^2\ \ \ .
\label{eq:factor1}
\end{equation}

On account of  a  requirement that $Z$ should be modular 
invariant, $|\lambda_1 \lambda_2|$ can be a function of only  
$V$ and $\s$ at most. 
There seems no further principle for fixing $|\lambda_1 
\lambda_2|$. 
Only when we choose as $|\lambda_1 \lambda_2|=V^{-2}$, 
the factor $\det{}^{-1/2} (\varphi_\a , \varphi_\b)$  
in  Eq.(\ref{eq:partition5}) or Eq.(\ref{eq:partition7}) has no 
influence. 
No such subtlety occurs in the string theory, since $\s$ does 
not appear and since $V$ is not important on account of the 
conformal invariance 
 [except for, of course, the conformal anomaly].

It is easy to see that, in our representation, 
 the reduced action in the $(\tau, V)$-form 
becomes 
\begin{eqnarray}
 S &=&\int_{t_1}^{t_2} dt\  
 (p_A \dot{\tau}^A + \s \dot{V} - N'(t) H ) \ \ \ , 
                                              \nonumber  \\
  H &=& {  (\tau^2)^2 \over {2V} }
   (p_1^2 + p_2^2)-{1\over 2} \s^2 V - \Lambda V\ \ \ .
 \label{eq:reducedaction} 
 \end{eqnarray}
 Here $\lambda=-\Lambda$ ($\Lambda>0$) corresponds to 
 the negative cosmological constant, which is set to zero when 
 it is not considered. 
 [The introduction of $\lambda$ ($<0$) may be preferable to 
 sidestep a subtlety of the existence of a special solution 
 $p_1=p_2=\s=0$ for  $\lambda=0$. This special solution forms a 
 conical singularity in the reduced phase space, which has been 
 already discussed in Ref.\cite{MONC} and in Ref.\cite{CAR2}.]
 Therefore, we get
 \begin{equation}
 - {\partial  H \over \partial V } = 
 {  (\tau^2)^2 \over {2V^2} }(p_1^2 + p_2^2)+{1\over 2} \s^2+ 
 \Lambda \ \ \ .
 \label{eq:factor2}
 \end{equation}
 
  As discussed in \S \ref{section:I},  the choice of 
  $N=$ spatially constant, which is consistent with 
  the equations of motion, is essential in the $(\tau,V)$-form. 
  This procedure can be  however influential  quantum 
  mechanically, so that its quantum theoretical effects should 
  be investigated. 
  In particular we need to understand the origin of 
  the factor $ {\partial  H \over \partial V }$ in 
  Eq.(\ref{eq:partition8}). 
  
 Let us start from the action in Eq.(\ref{eq:reducedaction}). 
It possesses a time reparameterization invariance:
\begin{eqnarray}
\d \tau^A = \e (t)\{ \tau^A, H  \}\ , 
                 &{}&  \d p_A = \e (t)\{ p_A , H  \}\ ,
                                          \nonumber \\
\d V = \e (t)\{ V, H  \}\ , &{}&  \d \s 
= \e (t)\{ \s , H  \}\ ,   
                                  \nonumber  \\
 \d N' = \dot{\e}(t)\   &{}&  {\rm with}\ \ \e (t_1)=\e (t_2)=0
 \ \ \ .      
\label{eq:reparameterization}
\end{eqnarray}
In order to quantize this system, one needs to fix a time
 variable. One possible gauge-fixing condition is 
 $\chi :=\s-t=0$. Then according
 to the Faddeev-Popov procedure, the factor 
 $\{ \chi, H \}\ \d (\chi) 
 = -{\partial  H \over \partial V }\  \d (\s-t)$ is inserted 
 into the formal expression for $Z$. The result is equivalent to 
 Eq.(\ref{eq:partition7b}) up to the factor 
 $\cal A$. 

Now we understand the origin of the nontrivial factor  
${\partial  H \over \partial V }$ in Eq.(\ref{eq:partition7b}). 
In order to shift from the $(\tau, V)$-form to the 
$\tau$-form, it is necessary to demote the virtual dynamical 
variables $V$ and $\s$ to the Hamiltonian and 
the time parameter, 
respectively. Then, the factor ${\partial  H \over \partial V }$ 
appears as  the Faddeev-Popov determinant associated with a 
particular time gauge  $\s=t$. 

In this manner, we found that the $(\tau, V)$-form is 
equivalent to the $\tau$-form even in the quantum theory,  
provided that the time-reparameterization symmetry remnant in 
the $(\tau, V)$-form is gauge-fixed by a particular 
condition $\chi:=\s-t=0$. In particular the key procedure of 
imposing $N=$spatially constant~\cite{HOS} turned out to be 
independent of the equations of motion themselves 
and valid in the quantum theory.
[Of course the fact that 
it does not contradict with the equations of motion is 
important.] 

Finally it is appropriate to mention the relation of the
 present result with the previous one obtained in 
 Ref.\cite{MS1}. In Ref.\cite{MS1} also,  the case 
 of $g=1$ was analyzed although the 
model was restricted to be spatially homogeneous
 in the beginning.  The result 
there was that the factor ${\partial  H \over \partial V }$ did 
not appear in the measure although the $(\tau,V)$-form was 
adopted. This result is reasonable because in Ref.\cite{MS1} 
only the spatial diffeomorphism symmetry associated 
with ${\cal H}^a$ was gauge-fixed 
explicitly. As for the symmetry associated with ${\cal H}$, 
the Dirac-Wheeler-DeWitt procedure was applied instead of the 
explicit gauge-fixing. 
[Alternatively, one can   regard that  the symmetry 
associated with ${\cal H}$ was fixed  by a non-canonical gauge
$\dot{N}=0$~\cite{HALL}.] Therefore it is reasonable that 
${\partial  H \over \partial V }$ did not appear in the 
analysis of Ref.\cite{MS1}. Thus the result of Ref.\cite{MS1} 
is compatible with the present result.


\section{Discussions}
\label{section:IV}

We have investigated how a partition function for 
(2+1)-dimensional pure Einstein gravity, formally defined in  
Eq.(\ref{eq:partition}), yields a partition function defined on 
a reduced phase space by gauge fixing.
We have shown that Eq.(\ref{eq:partition}) reduces to 
Eq.(\ref{eq:partition7}), which is interpreted as 
a partition function for a reduced system in the $\tau$-form. 
For the case of $g \geq  2$, this result is compatible 
with Carlip's analysis~\cite{CAR2}.

 For the case of $g=1$ with the option (\ref{item:b}), a factor 
$\det{}^{-1/2} (\varphi_\a , \varphi_\b)$ arises as a 
consequence  of the fact that $\dim \ker P_1 \neq 0$. 
This factor can be influential 
except when the choice 
$\det{}^{1/2}(\varphi_\a , \varphi_\b) =1$ is 
justified. The requirement of the modular invariance is not 
enough to fix  this factor.

Furthermore Eq.(\ref{eq:partition}) has turned out to reduce to  
Eq.(\ref{eq:partition8}), which is interpreted as a partition 
function
 for a reduced system in the $(\tau, V)$-form with a 
 nontrivial measure factor  
 $  {\partial  H \over \partial V }$ as well as  the possible 
 factor 
 $\det{}^{-1/2} (\varphi_\a , \varphi_\b)$. The former factor  
 was interpreted as the Faddeev-Popov determinant 
 associated with the time gauge $\s=t$, which was necessary 
 to convert  from the $(\tau, V)$-form to  the $\tau$-form.
  The choice of $N=$ spatially constant was
the essential element to derive the $(\tau, V)$-form in the 
classical theory. 
 In particular the   equations of motion were used to show 
its compatibility  with  York's gauge~\cite{HOS}. Therefore the 
relation of the $(\tau, V)$-form  with the $\tau$-form in the 
quantum level was required to be clarified. Moreover, 
since the condition $N=$ spatially constant is not in the 
form of the canonical gauge, the analysis of 
its role in the quantum level was intriguing.
Our analysis based on 
 the path-integral formalism turned out to be powerful 
 for studying  these issues.
Our result shows that 
 the $(\tau, V)$-form is 
equivalent to the $\tau$-form even in the quantum theory,  
as far as  the time-reparameterization symmetry  in 
the $(\tau, V)$-form is gauge-fixed by $\chi:=\s-t=0$. 
The postulation of  $N=$spatially constant in deriving the 
$(\tau, V)$-form turned out to be 
independent of the equations of motion  
and harmless even in the quantum theory.

 These results are quite suggestive to   quantum gravity and 
 quantum cosmology.

 First of all, the  measure factor similar to 
 $\det{}^{-1/2} (\varphi_\a , \varphi_\b)$ is  likely  to appear
  whenever a class of spatial geometries in question allows  
  conformal Killing vectors ($\ker P_1 \neq \emptyset$). 
  This factor can be influential on the semiclassical 
  behavior of the Universe. 
 
 The  issue of   the two options 
 (\ref{item:a}) and (\ref{item:b}) regarding the path-integral 
 domain of the shift vector (\S \ref{section:II}) 
 is interesting from 
a general viewpoint of gravitational systems. If one imposes 
that there should be no extra factor in the path-integral 
measure  for the reduced system, then the option 
 (\ref{item:a}) is preferred. There may be other arguments which 
 prefer one of the two options. 
 [For instance, general covariance of $Z$.] It may be interesting 
 if a similar  situation like  $\ker P_1 \neq 0$ occurred in the 
 asymptotic flat spacetime. In this case, the choice of the options 
 may be  influential  to  the gravitational momentum.
  
As another issue, the variety of representations of the same 
system in the classical level and the variety of 
the gauge-fixing conditions result in 
different quantum theories in general, and the relation
 between them should be more clarified. 
 The model analyzed here shall be a 
good test case for the study of this issue.

To summarize what we have learnt and to recognize  what is 
needed to be clarified  more, 
it is helpful to place our system beside a simpler system with a 
similar structure.  
The system of a relativistic particle~\cite{HART} is 
an appropriate model for illustrating the relation between 
the $\tau$-form and the $(\tau, V)$-form. 

Let $x^\a:=(x^0, \  \vec{x})$ and $p_\a:=(p_0, \  \vec{p})$ be 
the world point and the four momentum, respectively, 
 of a relativistic particle. Taking $x^0$ as the time parameter, 
 the action for the (positive energy) relativistic particle with 
 rest mass $m$ is 
 given by
 \begin{equation}
S=\int dx^0\  (\vec{p} \cdot {d\vec{x} \over dx^0} 
 - \sqrt{\vec{p}^2 + m^2 })\ \ \ . 
 \label{eq:particle1} 
\end{equation}  
Eq.(\ref{eq:particle1}) corresponds to the $\tau$-form 
(Eq.(\ref{eq:tau})). Now one can promote $x^0$ to a 
dynamical variable:
\begin{equation}
S=\int dt\  \{ p_\a  \dot{x}^\a 
 - N(p_0 + \sqrt{\vec{p}^2 + m^2 }) \}   \ \ \ . 
 \label{eq:particle2} 
\end{equation}
Here $t$ is an arbitrary parameter s.t. $x^0(t)$ becomes a 
monotonic function of $t$; $N$ 
is the Lagrange multiplier enforcing 
a constraint $p_0+ \sqrt{\vec{p}^2 + m^2 }=0$.
 The action  Eq.(\ref{eq:particle2}) corresponds 
to the action appearing in Eq.(\ref{eq:partition6}).  

It is possible to take  $p^2+m^2=0$ with $p_0 < 0$
  as a  constraint instead of 
$p_0+ \sqrt{\vec{p}^2 + m^2 }=0$.  Then an alternative action 
for the same system is given by 
\begin{eqnarray}
S&=&\int_{t_1}^{t_2} dt\  \{ p_\a  \dot{x}^\a - N\  H \}   \ \ \ ,
                                                   \nonumber \\
&{}& H=p^2+m^2 \ \ \ .       
 \label{eq:particle3} 
\end{eqnarray}
Eq.(\ref{eq:particle3}) corresponds to the $(\tau, V)$-form
 (Eq.(\ref{eq:tauV}) or Eq.(\ref{eq:reducedaction})). 
 
 The system defined by Eq.(\ref{eq:particle3})  possesses the 
 time reparameterization invariance similar to 
 Eq.(\ref{eq:reparameterization}): 
\begin{eqnarray}
\d x^\a = \e (t)\{ x^\a , H  \}\ , 
                 &{}&  \d p_\a = \e (t)\{ p_\a , H  \}\ ,
                                          \nonumber \\
\d N = \dot{\e}(t)\   &{}&  {\rm with}\ \ \e (t_1)=\e (t_2)=0
\ \ \ .    
\label{eq:reparameterization2}
\end{eqnarray}
Thus the gauge-fixing is needed in order to quantize this 
system. Here let us concentrate on two kinds of the
 gauge-fixing condition:
\begin{enumerate}
\item $\chi_{{}_I} := x^0 -t =0\ \ $ (canonical gauge),
\label{item:I}
\item $\chi_{{}_{II}} := \dot{N} =0\ \ $ (non-canonical gauge).
\label{item:II}
\end{enumerate}

 Choosing the gauge condition (\ref{item:I}), 
 one inserts the factors
$\{ \chi_{{}_I}, H \}$$ \d (\chi_{{}_I}) =-2p_0 $
$ \d (x^0 -t) $ into 
the path-integral formula according to the 
Faddeev-Popov procedure~\cite{FAD}. More rigorously, the factors
$\theta (-p_0) $$\{ \chi_{{}_I}, H \}\ $$ \d (\chi_{{}_I})$, 
 or alternatively, 
$\theta (N)  $$\{ \chi_{{}_I}, H \}\ $$ \d (\chi_{{}_I})$ 
should be inserted in order to obtain the equivalent
 quantum theory to the one obtained by 
 Eq.(\ref{eq:particle1})~\cite{HART}.
 The factor $\theta (-p_0)$ selects the positive energy solution 
 $-p_0 = \sqrt{\vec{p}^2 + m^2 }$ among the two solutions of 
 $H=p^2 +m^2=0$ w.r.t. $p_0$.
 This gauge (\ref{item:I}) corresponds to the gauge $\chi = \s-t=0$ 
 in the previous section. 
 We observe that a pair $(x^0,\ p_0)$ corresponds to 
 the pair $(\s,\  -V)$ 
 which is obtained from an original pair $(V,\ \s)$ by a simple 
 canonical transformation. 
 [The relation $-p_0=\sqrt{\vec{p}^2 + m^2 }$
  corresponds to the relation $V=V(\s, \tau^A, p_A)$.] 
  Thus the additional restriction factor 
 $\theta (-p_0)$ should correspond to $\theta (V)$, which is 
 identically unity because of the positivity of $V$. 
 It is quite suggestive that one solution among the two 
 solutions of $H=0$ (Eq.(\ref{eq:reducedaction})) w.r.t. $V$ 
 is automatically 
 selected because $V$ is the 2-volume of $\Sigma$.

As for the other gauge (\ref{item:II}) 
$\chi_{{}_{II}} := \dot{N} =0$, 
it is 
quite different in nature compared with  (\ref{item:I})
$\chi_{{}_I} := x^0 -t =0$. Apparently the path-integral 
measure becomes different. This point becomes clear if 
the transition amplitude $(\ x^\a_2\  |\  x^\a_1\ )$ for the
 system Eq.(\ref{eq:particle3}) is calculated by 
 imposing (\ref{item:I}) and 
by imposing (\ref{item:II}). 
By the canonical gauge (\ref{item:I}), one obtains 
\[
(\ x^\a_2\  |\  x^\a_1\ )_{I} = \int d^4p \  
        \exp \{ i p_\a (x^\a_2 - x^\a_1) \}\  |-2p^0|\ 
        \d (p^2 +m^2)\ \ \ ,
\]
if the simplest skeletonization scheme is adopted as in 
Ref.\cite{HART}. (Here we set aside the question about the 
equivalence with the system described by Eq.(\ref{eq:particle1}) 
so that the factor $\theta (-p^0)$ is not inserted.)
 The gauge (\ref{item:II}) can be handled~\cite{HALL} by 
 the Batalin-Fradkin-Vilkovisky method~\cite{BAT} 
 instead of the Faddeev-Popov 
 method, and the result is 
 \[
 (\ x^\a_2\  |\  x^\a_1\ )_{II} = \int d^4p \ 
        \exp \{ i p_\a (x^\a_2 - x^\a_1)\}\  \ \d (p^2 +m^2)
        \ \ \ .
\]
Both $(\ x^\a_2\  |\  x^\a_1\ )_{I}$ and 
$(\ x^\a_2\  |\  x^\a_1\ )_{II}$
 satisfy the Wheeler-DeWitt equation but they are clearly 
 different. One finds that if another gauge (\ref{item:I}') 
 $\chi_{{}_{I'}} := -{x^0 \over 2 p_0} -t =0$ is adopted 
 instead of 
 (\ref{item:I}), the resultant $(\ x^\a_2\  |\  x^\a_1\ )_{I'}$ is 
 equivalent to 
 $(\ x^\a_2\  |\  x^\a_1\ )_{II}$. One sees  that 
 $-{x^0 \over 2 p_0} \propto$$x^0 \sqrt{1- ({v \over c})^2}$
  under the condition $H=0$, which is interpreted as 
  the proper-time.

 Even in the present simple model, it is already clear that 
 only solving the Wheeler-DeWitt equation is not enough to 
 reveal  the quantum nature of the spacetime. 
 Then it is intriguing what the 
 relation there is  between the gauge conditions and the 
 boundary conditions for the Wheeler-DeWitt equation. 
 Apparently  more investigations are needed regarding 
  the gauge-fixing conditions, 
  especially the relation between the canonical gauges 
 and the noncanonical gauges.
 The system of (2+1)-dimensional Einstein gravity shall serve as 
 a good test candidate  to investigate  these points 
 in the context of quantum cosmology.

\begin{center}
\bf Acknowledgments
\end{center}

The author would like to thank H. Kodama and S. Mukohyama for 
valuable discussions. 
He also thanks S. Carlip for very helpful comments regarding 
the path integral over the shift vector. 
Part of this work was done during  the author's  stay at   
the Tufts Institute of Cosmology. He heartily thanks 
the colleagues there for providing him with a nice research 
environment.  
  
 This work was supported by the Yukawa Memorial Foundation, 
 the Japan Association for Mathematical 
Sciences,  and the Japan Society for the 
Promotion of Science.

\vskip 2cm

\makeatletter
\@addtoreset{equation}{section}
\def\theequation{\thesection\arabic{equation}}
\makeatother

\appendix

\begin{center}
\bf APPENDIX
\end{center}

Here we derive some   formulas that are  useful in our analysis.

\section{The Jacobian associated with change of 
integral variables}
We often need to change  integral variables in path integrals.
 Let $X^A$ and $X^{A'}$ ($A,A'= 1, \cdots, n$) 
 denote the original and the new variables, 
 respectively, in terms of which the line element is given as 
 $ds^2 = G_{AB}dX^A dX^B (=:(dX, dX))= G_{A'B'}dX^{A'} dX^{B'}$. 
 Then, a natural invariant measure becomes 
 $[dX]=d^nX \sqrt{\det  G}$
 $= d^nX' \sqrt{\det  G'}$. In other words,
  we define a measure in an invariant manner to satisfy 
  $1= \int[dX] \exp (-(\d X , \d X)) $.
 Now, a convenient way to find out the Jacobian $J$ associated 
 with the  
 change of variables, $X^A \rightarrow X^{A'}$, is~\cite{ALV}:
\begin{enumerate}
\item Represent $\d X^A$ in terms of $\d X^{A'}$, 
       $ \d X^A = {\partial X^A \over \partial X^{A'}} \d X^{A'}$.
\item Represent $(\d X, \d X)$ in terms of $\d X^{A'}$, 
         \[
         (\d X, \d X)= G_{AB} 
                   {\partial X^A \over \partial X^{A'}}
                   {\partial X^B \over \partial X^{B'}}
                  \d X^{A'}\d X^{B'}\ \ .
         \]
\item The Jacobian is given by setting 
\begin{equation}
1=J\int d^n \d X' \exp(-(\d X, \d X))\ \ \ ,
\label{eq:Jacobian}
\end{equation}
      since this should be equivalent to 
      $1=J (\sqrt{\det G'})^{-1}$ up to some  unimportant
    numerical factor.
\end{enumerate}

Here it may be appropriate to mention  the natural line element 
in our case. The `kinetic term' $K$ in the Hamiltonian 
constraint defines the geometrical structure of the 
configuration space. It is in the following 
 form (see Eq.(\ref{eq:hamiltonian})),  
 \[
 K= 
 \int_{\Sigma} d^2x \h \  
 (h_{ac}h_{bd} -h_{ab}h_{cd}){\pi^{ab}\over \h} 
 {\pi^{cd}\over \h}\ \ \ ,
 \]
 where $\Sigma$ stands for a 2-surface. Therefore,
  the inner product between 2nd-rank tensor fields which is 
  compatible with the geometrical structure of the configuration 
  space is given by 
  \[
  (w^{\cdot \cdot},\  {w'}^{\cdot \cdot})
    :=\int_{\Sigma} d^2x \h \ 
    (h_{ac}h_{bd} -h_{ab}h_{cd}) w^{ab} {w'}^{cd} \ \ \ .
  \]
  Furthermore, the second term in the parenthesis 
  is not important 
  in the following sense. 
   We observe  that 
\begin{eqnarray*}
  (w^{\cdot \cdot},\  {w'}^{\cdot \cdot})_k 
  :&=& \int_{\Sigma} d^2x \h \ 
               (h_{ac}h_{bd} + k\ h_{ab}h_{cd}) 
               w^{ab} {w'}^{cd} \\
   &=& \int_{\Sigma} d^2x \h \ \{ \tilde{w}^{ab} \tilde{w'}_{ab} 
         + (1/2 + k) ww' \}\ \ \ , 
\end{eqnarray*}
where $\tilde{w}^{ab}$ and $w$ stand for the traceless part 
and the trace of 
$w^{ab}$, respectively. Therefore, 
as far as the path integral 
concerned, the effect  of the value of $k$  is  absorbed 
into the normalization factor ${\cal N}$~\cite{HAT}. 
[We exclude the singular case $k=-1/2$.]\footnote
{The Euclidean path integral with $k < -1/2$ causes 
a trouble of divergence, which requires a special care. 
We shall not discuss this issue here and understand that 
 the Lorentzian path integral is adopted for such a case.} 
 Thus we simply set $k=0$. In this manner we are given  the 
 natural inner product between 2nd-rank tensor fields, which is 
 diffeomorphism invariant in accordance with 
 the principle of relativity. Afterwards we can extend the 
 inner product 
 to other types of fields also. For instance
\begin{eqnarray*}
(f,\ g):&=& \int_{\Sigma} d^2x \h\ fg \ \ ,  \\
(u^{\cdot},\  {u'}^{\cdot})
       :&=& \int_{\Sigma} d^2x \h\ h_{ab}u^a u^b \ \ , \\
(w^{\cdot \cdot},\  {w'}^{\cdot \cdot})
    :&=& \int_{\Sigma} d^2x \h \ h_{ac}h_{bd} 
    w^{ab} {w'}^{cd} \ \ \ .
\end{eqnarray*} 
For the case of densitized fields,  an appropriate  
power of $\h$ should be multiplied to the integrand in order 
  to make the inner product diffeomorphism invariant.

\section{A formula for the delta function}

Here we derive a formula which is essential in our  
discussions~\cite{MS1}.

Let $A$ be a linear (Fredholm) operator possibly with zero modes.
 Suppose we evaluate an integral
 $I= \int d^n X \d (A \vec{X}) f(\vec{X})$. 
 
   Let                                              
 $\{ {\vec \Psi}^A \}_{A=1, \cdots, \dim \ker A}$ be 
 a basis of $\ker A$. 
 Then   any vector $\vec X \in W$ in the 
 domain of $A$ can be uniquely decomposed as 
 ${\vec X} = {\vec X'} + p_A {\vec \Psi}^A$, where 
 $p_A= (X , \Psi^B ) {(\Psi^{\cdot}, \Psi^{\cdot})^{-1}}_{BA}$, 
 with  $(\Psi^{\cdot}, \Psi^{\cdot})^{-1}$ being the inverse 
 matrix of $(\Psi^A, \Psi^B)$. 
  Now, let us change the integral variables from 
  $\vec X$ to $({\vec X'} , p_A)$. The Jacobian $J$ for 
  this change is given as follows.  
  Noting that 
  $(\d X, \d  X) = (\d  X', \d  X')
   + (\Psi^A, \Psi^B) p_A p_B$, 
   we get 
   $J= \det{}^{1/2} (\Psi^A, \Psi^B)$ 
   (see Eq.(\ref{eq:Jacobian})).
   Then  $I$ can be expressed as 
 \[
  I=\int dp_A d{\vec X'}\  \det{}^{1/2} (\Psi^A, \Psi^B)\  
  \d (A {\vec X'})\  f({\vec X'}, p_A)\ .
  \]

We thus obtain a formula
\begin{equation}
\int d^n X \d (A \vec{X}) f(\vec{X}) 
= \int dp_A \   \det{}^{1/2} (\Psi^A, \Psi^B) (\det{}' A)^{-1}
   f({\vec X'}={\vec 0}, p_A)\ \ \ ,
 \label{eq:keyformula}
\end{equation}
 where $\det{}'A$ denotes the determinant of $A$ on 
 $W/ \ker A$ (i.e. the zero modes of $A$ are removed).


\end{document}